\documentclass[useAMS,usenatbib]{mn2e}
\usepackage{float,setspace,graphicx,amsmath,amssymb,perpage,pdflscape}

\MakeSorted{figure}
\MakeSorted{table}

\title[A catalogue of bright ($K<9$) M dwarfs]
  {A catalogue of bright ($K<9$) M dwarfs}
\author[J. Frith et al.]
  {J.~Frith,$^1$
  D. J. Pinfield,$^1$
  H.R.A. Jones$^1$
  J.R.Barnes,$^1$
  Y.Pavlenko,$^2$   
  E. L. Martin,$^3$
\newauthor
  C.Brown,$^4$
  M. K. Kuznetsov,$^3$
  F. Marocco,$^1$  
  R. Tata, $^3$ 
  M. Cappetta,$^5$
   \\
  $^1$ Centre for Astrophysics Research, Science and Technology Research Institute, University of Hertfordshire, Hatfield, Al10 9AB, UK\\
  $^2$ Main Astronomical Observatory of National Academy of Sciences of Ukraine, 27 Zabolotnoho, Kyiv 127, 03680, Ukraine\\
  $^3$ Centro de Astrobiolog ́ıa (INTA-CSIC), Carretera Ajalvir km 4, E-28850 Madrid, Spain\\
  $^4$ EADS Astrium, Gunnels Wood Rd, Stevenage, SG1 2AS, UK\\
  $^5$ Max-Planck-Institut fur extraterrestrische Physik, Giessenbachstraße, D-85748 Garching, Germany}
\date{Released 2012 Xxxxx XX}

\pagerange{\pageref{firstpage}--\pageref{lastpage}} \pubyear{2002}

\def\LaTeX{L\kern-.36em\raise.3ex\hbox{a}\kern-.15em
    T\kern-.1667em\lower.7ex\hbox{E}\kern-.125emX}

\begin{document}

\label{firstpage}

\maketitle

\begin{abstract}
Using the Position and Proper Motion Extended-L (PPMXL) catalogue, we have used optical and near-infrared colour cuts together with a reduced proper motion cut to find bright M dwarfs for future exoplanet transit studies.  PPMXL's low proper motion uncertainties allow us to probe down to smaller proper motions than previous similar studies.  We have combined unique objects found with this method to that of previous work to produce 8479 $K<9$ M dwarfs.  Low resolution spectroscopy was obtained of a sample of the objects found using this selection method to gain statistics on their spectral type and physical properties.  Results show a spectral type range of K7-M4V.  This catalogue is the most complete collection of $K<9$ M dwarfs currently available and is made available here.  
\end{abstract}


\section{Introduction}
	
	Identification of M dwarfs in large surveys has been difficult even though they are the most numerous type of stars in the Galaxy.  This is due to their colours and low luminosities often causing them to be confused with reddened stars or distant red giants.  M dwarfs have become of particular interest in recent years for their applications to exoplanet research. Their low masses and small radii lead to greater sensitivity to discovery of orbiting low mass planets via the radial velocity and transit techniques.  Furthermore, due to an M dwarf's low luminosity, planets lying in the habitable zone around M dwarfs have short orbital periods as well as relatively favourable contrast ratios between the planet and the host star.  New space facilities like the James Web Space Telescope \citep{JWST_exoplanets} (JWST) and the proposed Exoplanet Characterization Observatory \citep{echo} offer the potential to study exoplanet atmospheres for super Earths transiting M dwarfs but currently M dwarf transit detections are very rare.
	
5 M dwarfs are currently known to host transiting planets: GJ1214b, GJ3470, GJ436, KOI-961, and KOI-254. With the exception of the two Kepler systems \citep{Johnson2012,koi961}, the host stars are all bright in the near-infrared (K=6.1-8.8) and the systems are thus good targets for characterisation. Two of these were discovered using the radial velocity (RV) method (GJ436b and GJ3470b; \cite{Gillon2007}; \cite{Bonfils2012}) and have masses similar to Neptune (15-20$M_{\oplus}$), while GJ1214b was discovered by the Mearth survey using the transit method \citep{Charb2009} and has a mass of 5.9$M_{\oplus}$.  GJ1214b is currently the only known system with a Super-Earth transiting a bright M dwarf.  The first step towards expanding this number is to identify all potential bright M dwarf host stars in the sky for future transit search campaigns.

Other recent studies in this area (overlapping with this work) include the catalogue of L\'epine \& Gaidos (2011;LG11).  This sample was created from the LSPM catalogue \citep{superblink} and supplemented with Tycho-2 \citep{tycho2} objects but is limited at low proper motions due to the SUPERBLINK proper motion limit of 40 milli arcseconds (mas) per year.  Our efforts attempt to optimize the search at the low proper motion extreme and thus expand on the existing sample.  We have chosen to use the Position and Proper Motion Extended-L (PPMXL) \citep{ppmxl} as the starting point for a selection process that utilizes a multi-band colour selection and a reduced proper motion cut to isolate bright M dwarfs down to the lowest practical proper motion limits.

\section{Sample Selection}

\subsection{The PPMXL Catalogue}
The PPMXL catalogue represents a combination of the USNO-B1.0 \citep{USNO} and 2MASS \citep{2mass} catalogues mapped onto the International Celestial Reference Frame (ICRF) \citep{IRCS}.  The ICRF is based on distant extragalactic radio sources observed by the Very Large Baseline Array as its reference points which allows proper motions to be described in a quasi-absolute manner as opposed to relative.  Until the PPMXL catalogue was developed, only relative proper motions were available for the USNO-B1.0 objects and 2MASS did not have any proper motion information.  PPMXL now provides low uncertainties for both the proper motion and position for many of the objects within the 2 catalogues.  Typical uncertainties for proper motions are 4-10 mas/yr.  The near-infrared (NIR) JHK magnitudes from 2MASS and the optical BVRI magnitudes from USNO-B1.0 also provides very useful colour information about the objects and is used during the sample selection process.  At low proper motions, it is important to minimize measurement uncertainties so absolute proper motions will lead to less contamination as the astrometry becomes more noisy.

\subsection{Magnitude Limit}

A principal aim of this work is to target a stellar sample around which potential exoplanet atmospheres could be studied in the future.  It was therefore appropriate to limit the selection to sufficiently bright targets that provide enough signal to probe the molecules of interest in exoplanet atmospheres.  Even with bright host stars and favourable contrast ratios between the exoplanet and star, several observations of the transiting planet through both primary and secondary transits have to be co-added together to produce observations of any accuracy.  Because of this, integration time can be equated to the number of transits necessary to achieve the needed signal to noise.  Simulations were conducted by \cite{marcell} to estimate the number of transits needed for a range of exoplanet sizes and host star brightnesses. They have shown that for a space-based multi-channel spectrograph with a modest sized primary (1.2m) to achieve the signal to noise necessary to identify molecules in a transiting super earth in the habitable zone, the number of transits needed around M dwarfs with a magnitude of $K>9$ becomes unrealistic when planning a space mission observing schedule.  Therefore, a magnitude limit of $K<9$ was implemented.  Objects with uncertainties in JHK photometry greater than 0.05 magnitudes were also removed to limit saturated targets within the sample. 

\subsection{Guiding Samples}
\label{sec:guidsamp}
To help in our various colour and reduced proper motion selections, we identified several guiding samples using existing catalogues.  Several M dwarf samples exist in the literature that have been observed spectroscopically and photometrically across many bands.  In total, three M dwarf samples were used.  One sample was produced by \cite{bochanski} in a spectroscopic survey for cool stars within 100 parsecs of the Sun.  These M dwarfs were selected using one optical (R-I) colour cut with no proper motion selection and later confirmed with spectroscopy.  Another sample used was the Palomar/MSU (PMSU) spectroscopic survey \citep{PMSU} which is based on M dwarfs that were selected by using parallax measurements.  Spectra were obtained of objects found in the Third Catalogue of Nearby Stars (CNS3) \citep{cns3} which consists of stars within 25 parsecs of the Sun.  The third M dwarf selection was done by \cite{riaz} and was compiled to find active M dwarfs using X-ray observations from ROSAT.  This had limited NIR colour selection and no kinematic bias.  By using the combination of all 3 catalogues as guides, it is hoped that we can minimize any kinematic or photometric bias that may be incurred during our selection process.  All 3 catalogues were cross-matched with PPMXL with a cone radius of 10 arcseconds and the best matched objects used as comparisons to our sample.
	
Distant M giants were a major source of potential contamination so a guiding sample of giants was also used from \cite{Famaey}.  This sample consists of giants with known distances and absolute magnitudes from the Hipparcos catalogue \citep{hip}.  Though the majority of the sample consists of K giants, there are a significant number of M giants identified that were used below when attempting to remove potential giant contamination.  These objects were also cross matched with PPMXL so that all guiding samples had proper motion estimates mapped onto the ICRS.  

To supplement the M giant guiding samples, we constructed a photometric selection using colour cuts.  Our colour defined giant sample was created using a near infrared (NIR) colour selection from \cite{sharma} which is reproduced here:

\begin{align*}
K <14\\
J-K \geq 0.85\\ 
J-H < 0.561(J-K) + 0.36\\
J-H > 0.561(J-K) + 0.19\\
\end{align*}

\begin{figure*}
\includegraphics[scale=.33]{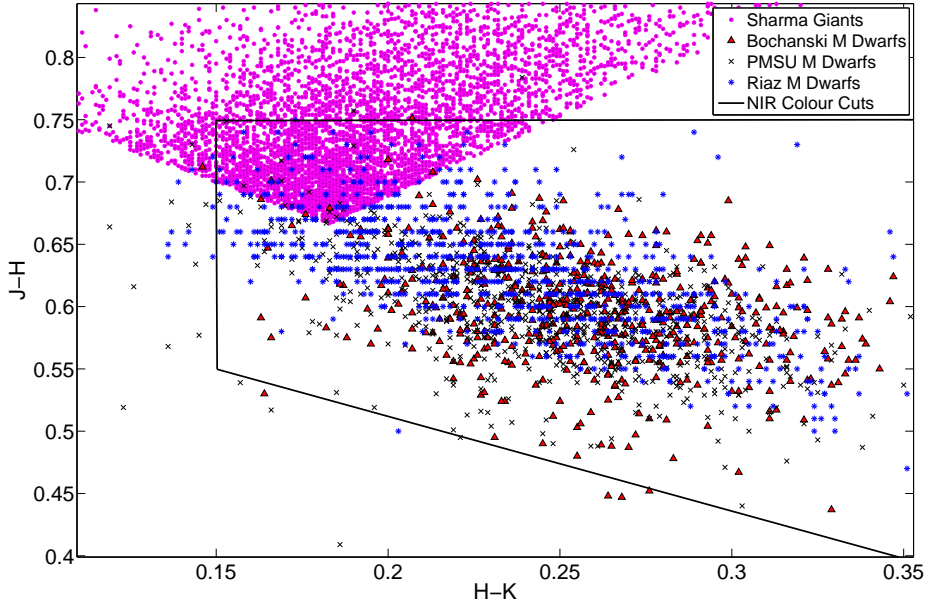}
	\centering
	\caption{Colour-colour plot showing the NIR selection and the M giant and M dwarf guiding samples.  The different coloured triangles are known M dwarfs from previous surveys and are used to illustrate where M dwarfs can be expected to be found.  Our colour defined M giant population created from PPMXL objects is shown as open squares.  The overlap between the two populations demonstrates that M giants can overlap the same colour space as dwarfs and shows that futher cuts are needed.}
	\label{fig:sampleselection_colour}
\end{figure*}

This selection was applied to areas of sky avoiding the galactic plane to reduce the number of reddened stars selected.  As seen in Figure \ref{fig:sampleselection_colour}, these cuts slightly overlap the region of M dwarfs in NIR colour space but the majority of the sample is well separated from the M dwarf region (described in the next section).  This overlap illustrates how colour cuts alone don't completely isolate an M dwarf or M giant population. 

\subsection{Colour Selections}	
	
	The initial colour selection for our sample was done using the 2MASS near-infrared JHK bands.  These cuts were made to separate the M dwarfs from M giants as well as to limit K dwarf contamination.  The guiding samples were used to define the colour regions where most M dwarfs lie as can be seen in Figure ~\ref{fig:sampleselection_colour}.  Our resulting NIR selection criteria for M dwarfs are
	 
\begin{align*}
J - K > 0.7\\
H­ - K > 0.15\\
J­ - H < 0.75
\end{align*}

\begin{figure}
\includegraphics[scale=.3]{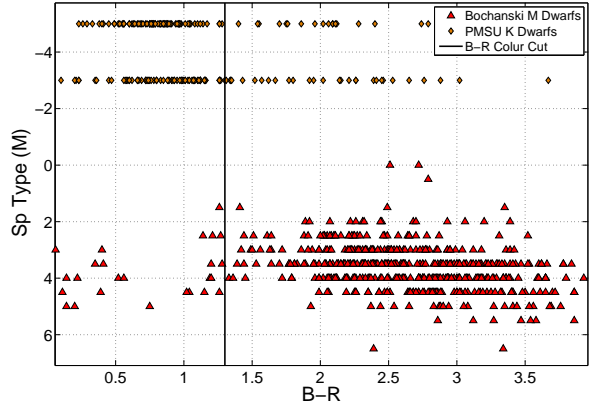}
	\centering
	\caption{Known M dwarfs plotted with PPMXL colours (red triangles) along with known K dwarfs (orange diamonds).  The optical cut (vertical black line) was chosen to optimize the inclusion of M dwarfs and to minimize K dwarf contamination by removing all objects bluer than a B-R of 1.3.  This process was repeated in all of the available optical pass bands in order to determine the optical colour cuts.}
	\label{fig:BminusR}
\end{figure}

These regions isolate the general area of the M dwarf population.  However, PPMXL also provides the photometry from USNO-B1 optical bands for many of its targets.  Spectroscopically confirmed M dwarfs in PPMXL combined with K dwarfs from PMSU were used to determine where K dwarfs separate from M dwarfs in the colour space.  All of the available colours were plotted against their spectral type and an example is seen in Figure ~\ref{fig:BminusR}.  For the colours where the K dwarfs weren't available, the cut was made on the blue end of the main body of the guiding samples.  The following selections are used to isolate M dwarfs.

\begin{align*}
B - ­R > 1.3\\
B - ­I > 2.5\\
V - J > 2.7\\
R - I > 0.9\\
I­ - J > 0.6\\
\end{align*}

	Applying all of the NIR and optical colour cuts to the existing sample from PMSU removes approximately 88\% of the K dwarfs and keeps 95\% of the M dwarfs which gives an indication of the amount of K dwarf contamination expected in the final selection of M dwarfs.

\subsection{Proper Motion Uncertainty}

Objects with high proper motion uncertainties were removed using the relationship 

\begin{align*}
	\mu < 5 \sigma	
\end{align*}	 

where $\mu$ is the total measured proper motion of the object and $\sigma$ is the listed uncertainty.  Doing this removed any objects which had uncertainties that were a significant fraction of their total proper motions. This obviously tended to remove objects that had very low proper motions ($\mu<15 mas$) or non-moving objects and helped to minimize giant contamination.

\subsection{Galactic Plane Removal}
\begin{figure}
\begin{center}
\includegraphics[scale=.4]{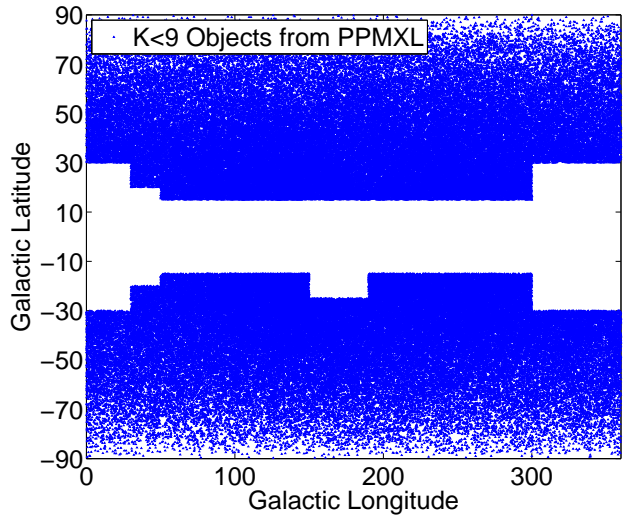}
	\caption{This shows PPMXL K$<$9 objects across the sky. Regions where we detected an over density of red stars were removed.  This roughly corresponded to areas of sky with a density above 45 stars per square degree.  The majority of the regions cut can be defined by simply removing anything within 15 degrees of the galactic plane.  Slightly higher galactic latitude regions near the galactic center where the density of stars was high (notably around a galactic longitude of 150-200) were also removed where there was an overdensity of very red objects.}
	\label{fig:gal_plane}
\end{center}
\end{figure}

Due to overcrowding and dust, stars within the plane of the Milky Way can often have unreliable colours and proper motions.  These objects were avoided by removing regions within the galactic plane that had a high density of stars and dust.  These regions can be seen in Figure~\ref{fig:gal_plane}.  After the initial K$<$9 cut, regions were removed where there was visible crowding or where there was an over density of red stars.

\subsection{Reduced Proper Motion Selection}

When distance to an object is not known, the reduced proper motion, H, can be used as a proxy for absolute magnitude.  H is defined as
	\begin{align*}
	H = m +5log \mu+5,
	\end{align*}
where m is the observed magnitude and $\mu$ is the proper motion in arcseconds/year.  This can also be used to separate different kinematic populations between the disk and halo but the primary use within this work is to separate the dwarfs from the giants within our sample.  Because our brightness limit is defined using K magnitudes, we chose to use this band to determine our reduced proper motion cuts.

\begin{figure*}
\begin{center}
\includegraphics[scale=.27]{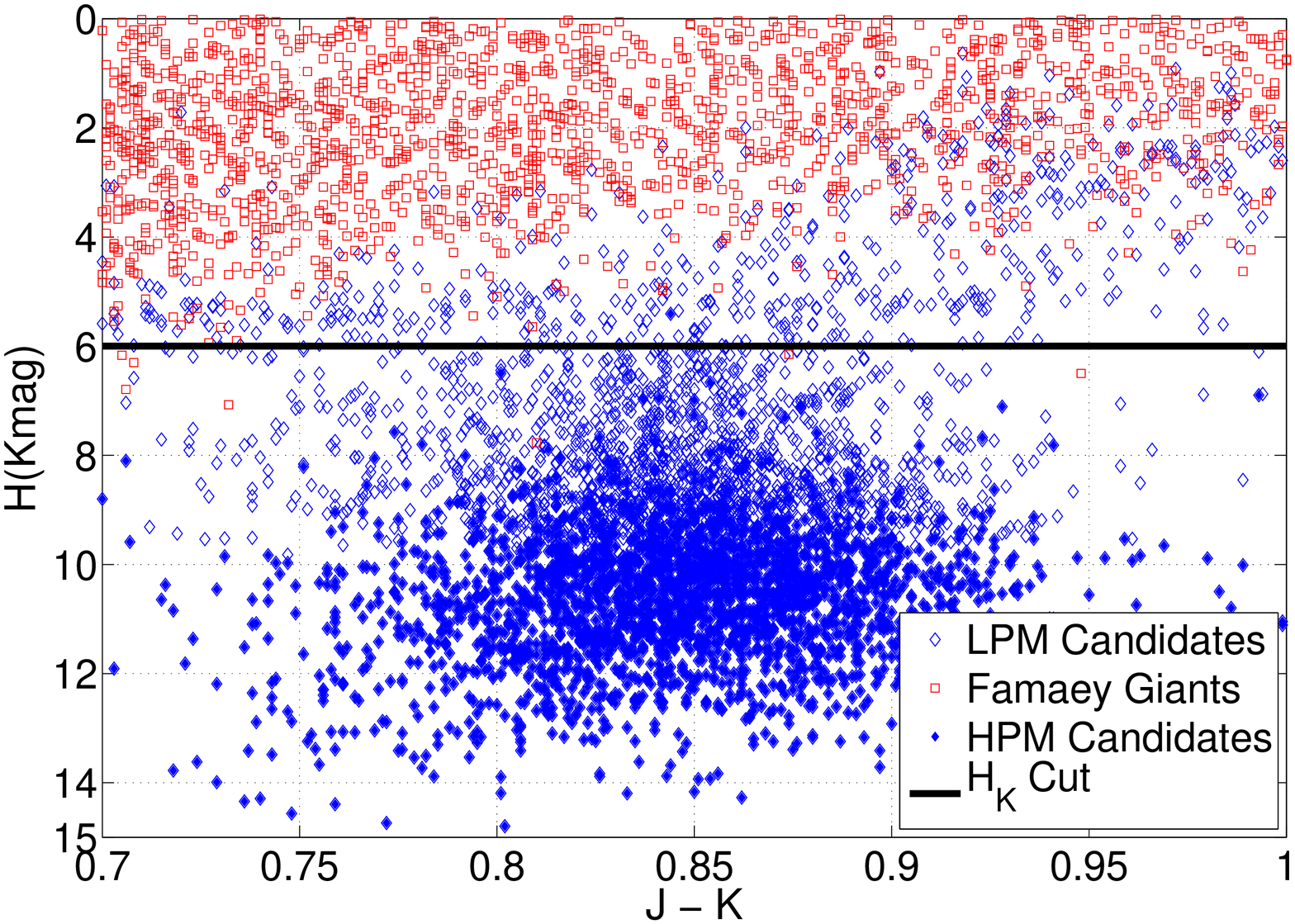}
\includegraphics[scale=.27]{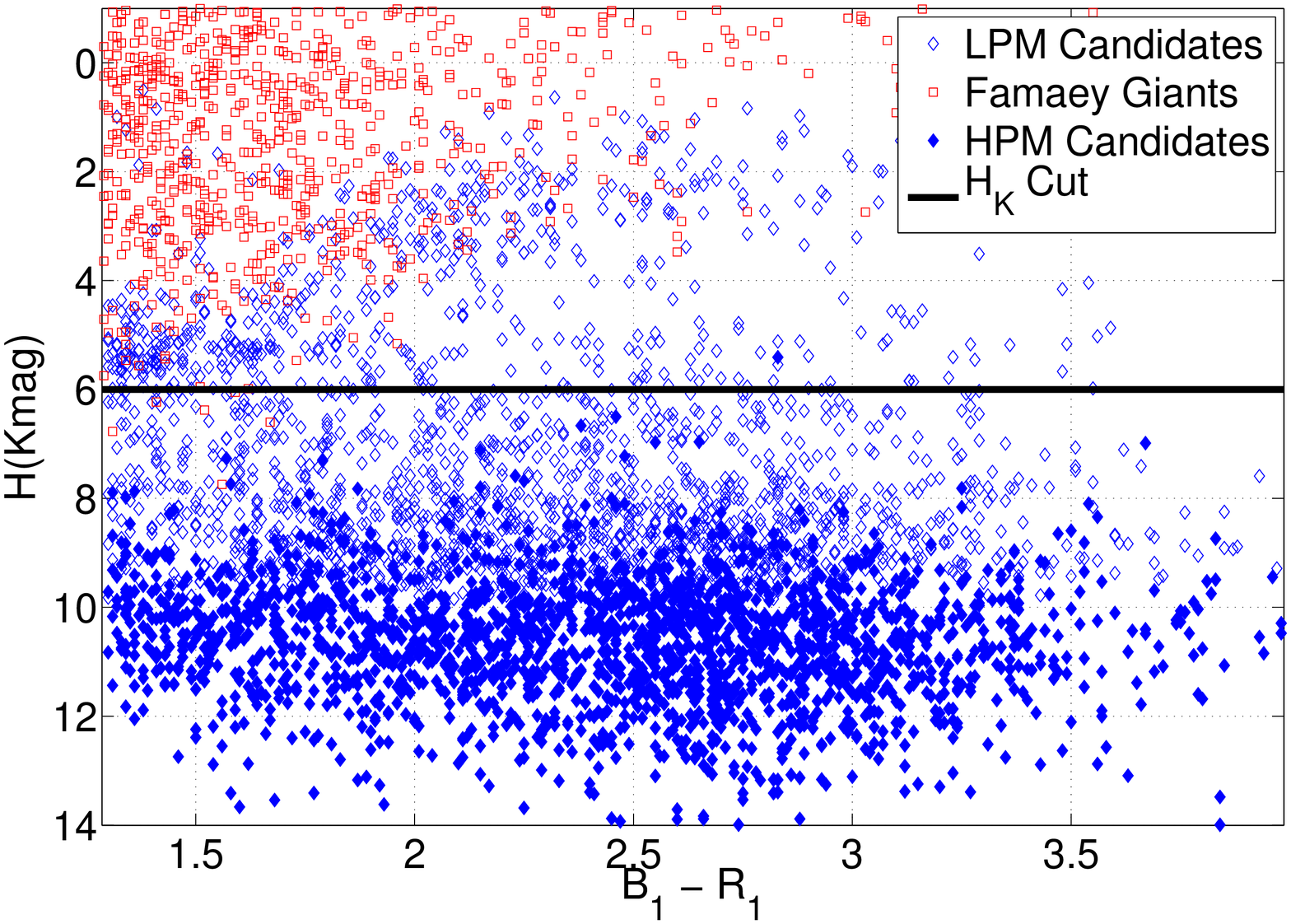}
\includegraphics[scale=.27]{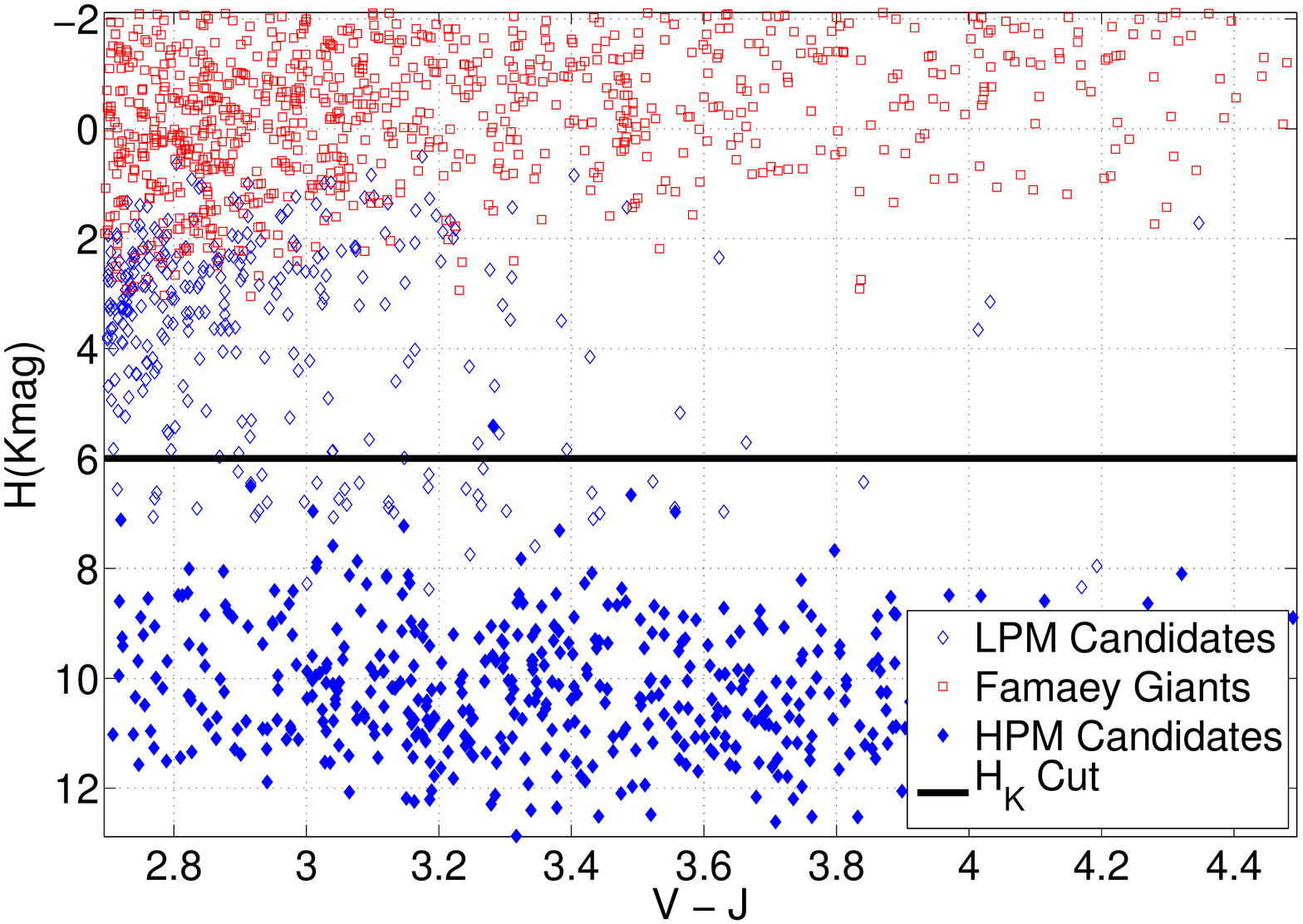}
\includegraphics[scale=.27]{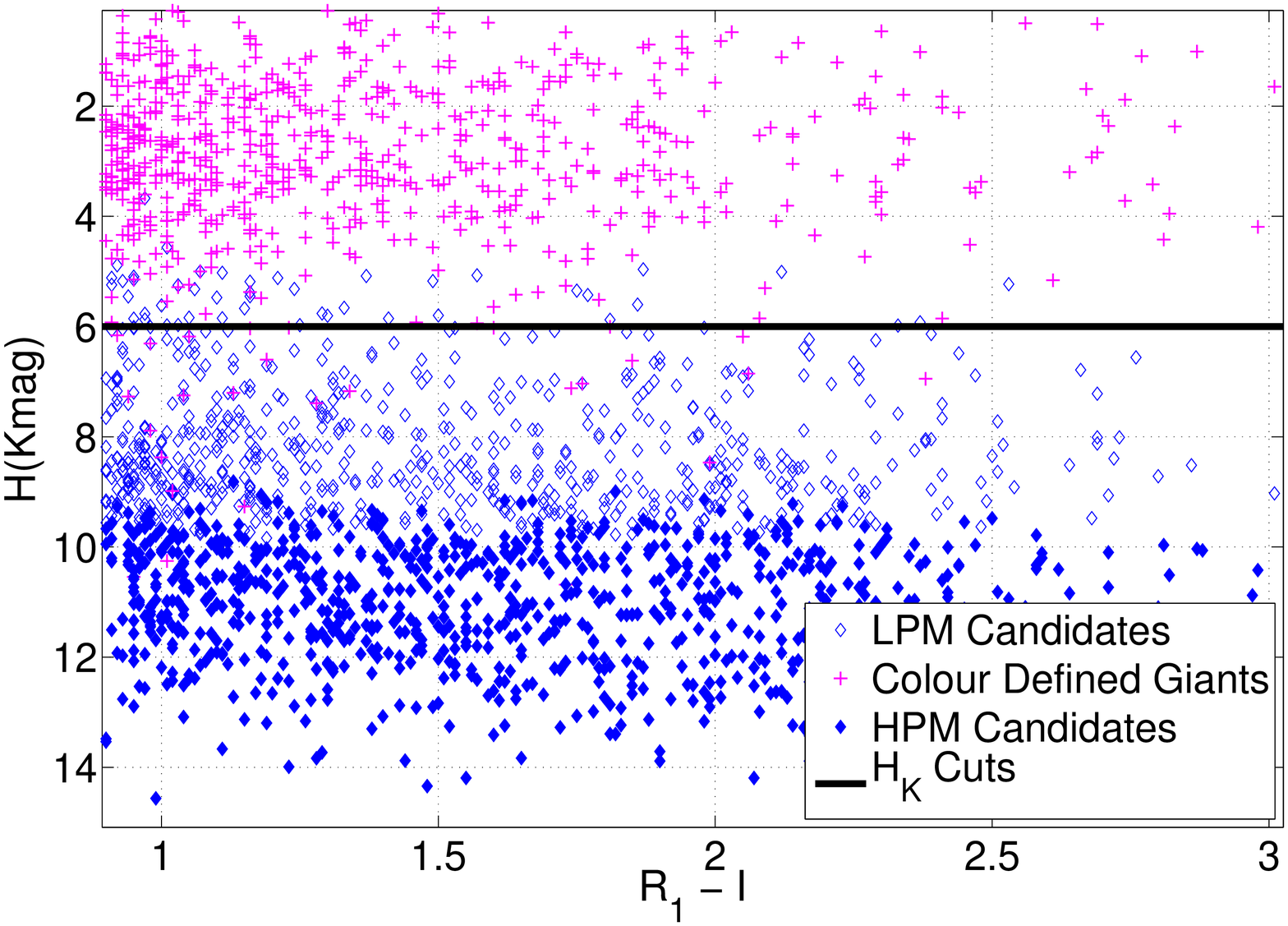}
\includegraphics[scale=.27]{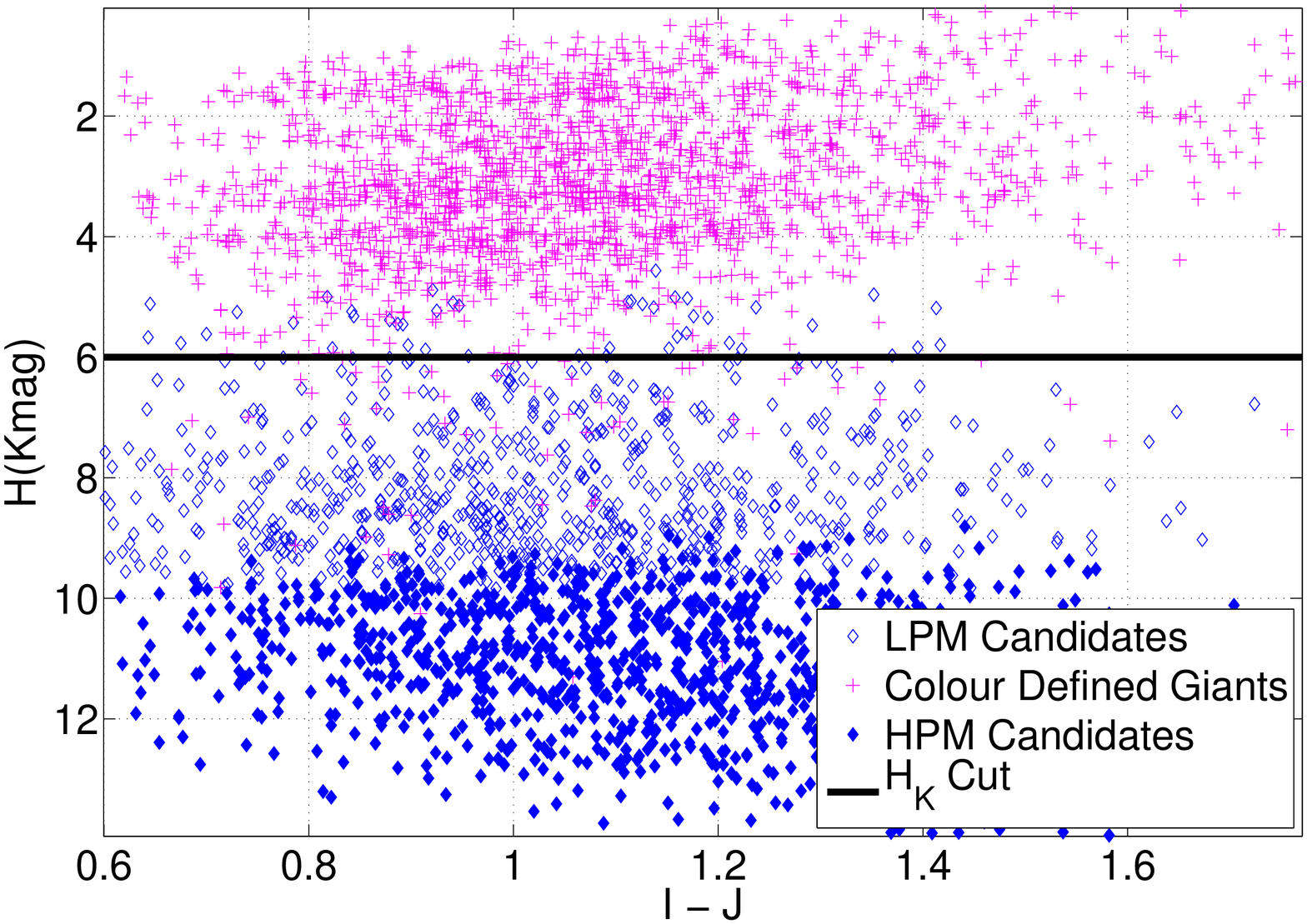}
	\caption{Reduced proper motion vs the multiple colour indices available for the M dwarf candidates.  The lines show the reduced proper motion cut used to remove the M giants from the sample.  High proper motion candidates (HPM) with $\mu>$150 mas/year from PPMXL have been plotted separately to better show the region where M dwarfs can be expected to lie since these objects are less likely to be mistaken for giants.  Known giants have been taken from the samples described in Section 2.3 to illustrate where giants can be found.  The Famaey giants did not have I magnitudes available so for the colours where I magnitude was needed, giants were defined using colour cuts from Sharma et al.  A reduced proper motion cut of $H_{k}\geq$ 6 was found to be high enough to safely remove objects that were likely to be giants and still low enough to include many low proper motion M dwarfs that were previously unclassified.}
	\label{fig:rpmcuts}
\end{center}
\end{figure*}

Since the reduced proper motion is analogous to absolute magnitude, plotting H along with a colour index provides us with a pseudo-equivalent of an Hertzsprung-Russell diagram where the brighter giant branch is separated vertically from the fainter red dwarfs.  We use this, along with the giant and M dwarf guide samples (discussed in section \ref{sec:guidsamp}), as a way of identifying the different populations within our sample.  We were able to take advantage of PPMXL's wide range of colour space to precisely explore the completeness of the chosen H cut and this is illustrated Figure~\ref{fig:rpmcuts}.  These plots show the PPMXL objects that remain after the NIR and optical colour cuts as blue diamonds. The high proper motion ($\mu > $150 mas/year) stars are plotted separately as filled in diamonds while those with $\mu <$150 mas/year as empty diamonds.  Giants from our guiding samples are plotted as red squares and magenta crosses.  We can see a clear vertical separation between the giant populations and the majority of the our objects.  The high proper motion objects extend in a clear manner down to an $H_{\rm K}=$ 6.0.  M dwarf candidates below $H_{\rm K}=$ 6.0 begin to become mixed with the two giant populations and over densities in the PPMXL objects can be seen in the bluer regions of the B-R and V-J plots around an $H_{\rm K}$ of 5.8.  After these considerations, objects with $H_{\rm K}\geq$ 6.0 were chosen to best optimize giant removal and dwarf selection.  This cut removed 293 objects.

To examine the completeness resulting from our reduced proper motion cut, we applied the same cut to our guiding samples as well.  The results of this can be seen in Table \ref{tab:completeness}.  The Bochanski and PMSU M dwarfs only lose about 4\% of their objects after this cut.  Both catalogues are proper motion independent and target stars using an optical colour cut so a similar completeness level between the two is to be expected.  However, almost 20\% are lost from the Riaz M dwarfs.  This is not surprising since the Riaz M dwarfs were selected primarily due to X-Ray activity and therefore have no lower proper motion cut off in their selection.  Because of this, many of its targets have proper motions that lie well below the threshold of our reduced proper motion cut.  From this, an incompleteness between 4 - 20\% is a reasonable estimate for our selection methods. Encouragingly, 98\% of our example giant star population is rejected using this reduced proper motion cut.

An $H_{k}\geq$ 6.0 cut corresponds to a lower proper motion limit for the faintest objects of 25 mas/yr. This allows us to probe down below the LG11 limit of 40 mas/yr for for targets fainter than K = 8.0.

\begin{table}
\caption{This table shows how many objects are removed from the 4 other guiding catalogues (cross matched with PPMXL) if an $H_{k}\geq$6.0 cut is applied.}
\label{tab:completeness}
\begin{tabular}{c | c}
\hline
Catalogue & Percent Rejected by $H_{k}\geq$6.0 \\
\hline
Bochanski M Dwarfs & 3.7\% \\
PMSU M Dwarfs & 4.0\% \\
Riaz M Dwarfs & 18.7\% \\
Famaey Giants & 98.7\% \\
\hline
\end{tabular}
\end{table}

\subsection{Tycho-2}
		
	For completeness, we have used the Tycho-2 catalogue \citep{tycho2} to augment our existing PPMXL sample.  The above NIR colour cuts were applied to 2MASS objects that shared a Tycho-2 detection and the reduced proper motion cut applied using proper motions from Tycho-2.  The resulting objects were crossmatched with PPMXL using a 5 arcsec radius and the other cuts were applied.  Many of the resulting objects were already included in the sample from PPMXL, however 93 of the objects were found to be unique and were added to the total catalogue.

\section{Final Sample}

After application of all the above cuts, 4054 bright M dwarf candidates remain.  In Table \ref{tab:Mdwarfnumbers}, we list the number of M dwarf candidates that remain across different K band magnitude ranges.  Many of these candidates fall into the same parameter space as LG11   so we have also listed objects from their catalogue.  A cross match was done between the two at a separation of 5 arcseconds and 2861 of the candidates were found to be shared.  The common object's are also included in Tabel \ref{tab:Mdwarfnumbers} for comparison.

\begin{figure}
\includegraphics[scale=.28]{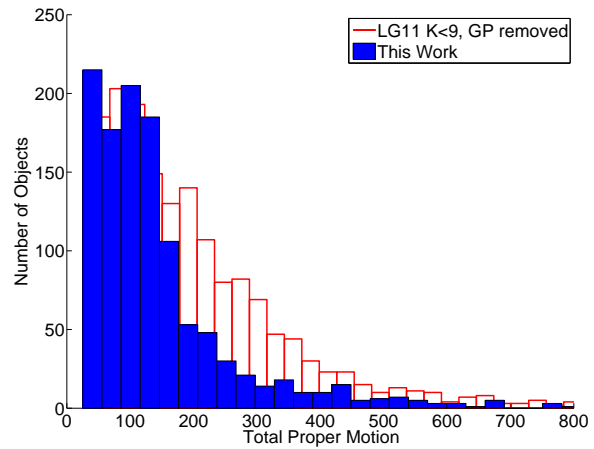}
	\centering
	\caption{Histogram showing the proper motion distribution of the objects that were unique to each sample.  The LG11 has had the same galactic plane cut as this work and a magnitude cut of K$<9$ to better compare the selection methods. LG11 is better on the high proper motion end whereas this work does slightly better at the low proper motion objects.}
	\label{fig:unique_hist}
\end{figure}

\begin{figure}
\includegraphics[scale=.25]{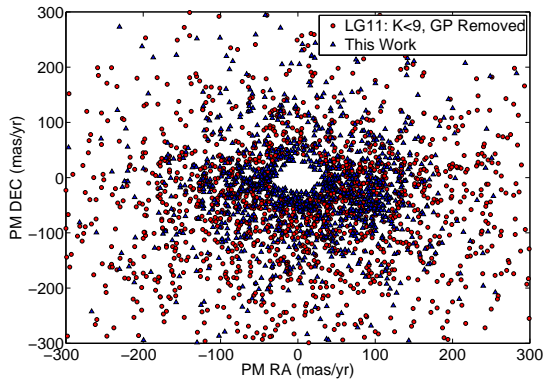}
	\centering
	\caption{Proper motion plot showing the unique objects from each catalogue and the low proper motion objects this work identifies that lie below 40 mas.}
	\label{fig:pm_unique}
\end{figure}

The two catalogues targeted slightly different magnitude ranges (K$<$9 and J$<$10 for this work and LG11 respectively) so to facilitate a comparison between this work and the LG11 sample we have applied two simple cuts to the catalogue of LG11.  We only select objects with K$<$9 and reject objects from the region of the galactic plane not included in our selection.  In this way, we can focus on the main differences between the two catalogues: colour selection, reduced proper motion selection, and using PPMXL or SUPERBLINK for the proper motion measurements.  As seen in Figures \ref{fig:unique_hist} and \ref{fig:pm_unique}, this work's sample is able to identify M dwarf candidates down to a lower proper motion whereas LG11 does better at higher proper motions.  This is expected, however, because PPMXL objects are cross matched between USNO-B1 and 2MASS with a cone radius of 2 arcseconds whereas the SUPERBLINK comparisons can be as large as 1.5 arcminuts.  This difference gives SUPERBLINK greater sensitivity to objects with higher proper motions.

\begin{figure}
\includegraphics[scale=.27]{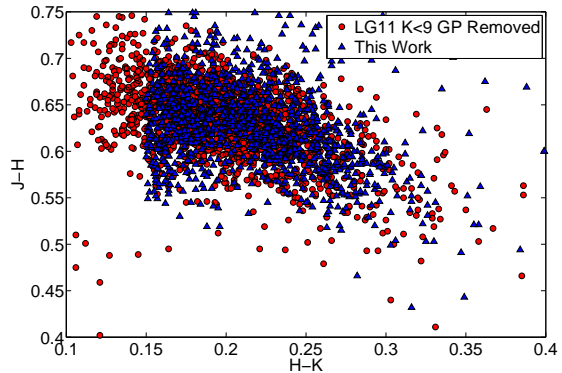}
	\centering
	\caption{Colour-colour diagram with the PPMXL selected M dwarfs as well as M dwarfs from LG11 (with magnitude and galactic plane cuts applied) to compare the two M dwarf selection methods.  The LG11 sample's cut in H-K space allowed slightly more blue objects into that selection than this work.}
	\label{fig:color_color_comp}
\end{figure}

Figure \ref{fig:color_color_comp} show's this work and LG11 in NIR colour space.  This shows that many of the objects in LG11 lie in a region that is bluer than the scope of this work and thus includes stars of earlier spectral types.

A machine readable file has been created containing all of the M dwarfs in this work as well as those that were found to be unique to LG11 in the $K<9$ magnitude range.  This file is thus the most complete sample in the $K<9$ brightness range currently available.  An example of entries in the catalogue can be seen in Table \ref{tab:catalog}.

\begin{table*}
\begin{minipage}[tbh]{1\linewidth}
\caption{Example of final bright M dwarf catalogue.}
\centering
\label{tab:catalog}
\begin{tabular}{@{}ccc@{}c@{}c@{}c@{}ccccccc@{}}
\hline
 $Index$ & $\alpha$ & $\delta$ & $\mu_{\alpha}$ & $\mu_{\delta}$ & $\sigma \mu_{\alpha}$ & $\sigma \mu_{\delta}$ & $K$\footnote{JHK magnitudes taken from 2MASS (mean uncertainty $\sim$0.02).} & $J-K$ & $H-K$ & $B$\footnote{B and R magnitudes taken from USNO-B1 (mean uncertainty $\sim$0.3 magnitudes).} & $R$ & $Fl$\footnote{Flag identifying origin of proper motion information(1:PPMXL, 2:SUPERBLINK, 3:Tycho-2).} \\
 & ($^{\circ}$) & ($^{\circ}$) & (mas $yr^-1$) & (mas $yr^-1$) & (mas $yr^-1$) & (mas $yr^-1$) & (mag) & (mag) & (mag) & (mag) & (mag) &  \\
\hline
FR0001 & 0.028530 & 69.717120 & 136.0 & -2.0 & -- & -- & 8.84 & 0.86 & 0.28 & -- & 12.60 & 2 \\
FR0002 & 0.087700 & -8.037150 & 29.0 & -96.0 & -- & -- & 8.27 & 0.85 & 0.20 & -- & 11.80 & 2 \\
FR0003 & 0.144917 & -5.552002 & 187.8 & 67.5 & 5.3 & 5.3 & 8.17 & 0.83 & 0.22 & 13.50 & 11.16 & 1 \\
FR0004 & 0.163530 & 18.488850 & 335.0 & 195.0 & -- & -- & 7.64 & 0.80 & 0.15 & 12.91 & 10.30 & 2\\
FR0005 & 0.195287 & -35.168330 & 355.6 & -114.9 & 4.1 & 4.1 & 8.28 & 0.84 & 0.20 & 13.20 & 10.86 & 1 \\
FR0006 & 0.195902 & 16.402781 & 12.4 & -135.1 & 4.8 & 4.8 & 8.46 & 0.86 & 0.23 & 14.40 & 11.38 & 1\\
FR0007 & 0.303580 & 13.972050 & 25.0 & 144.0 & -- & -- & 7.53 & 0.83 & 0.18 & 13.67 & -- & 2 \\
FR0008 & 0.357630 & -16.948410 & 299.0 & -255.0 & -- & -- & 7.22 & 0.80 & 0.19 & 12.14 & 9.80 & 2 \\
FR0009 & 0.371550 & 47.414660 & 170.0 & -4.0 & -- & -- & 8.83 & 0.84 & 0.19 & -- & 11.40 & 2 \\
FR0010 & 0.399370 & -8.244880 & 97.0 & -77.0 & -- & -- & 8.91 & 0.88 & 0.21 & -- & 11.70 & 2 \\
\hline
\end{tabular}
\textsl{This table is available in its entirety in machine readable format in the online journal.}
\end{minipage}
\end{table*}

\begin{table}
\centering
\caption{Number of bright M dwarfs that have been found by this work and LG11.}
\label{tab:Mdwarfnumbers}
\begin{tabular}{ccccc}
\hline
K & This Work & LG11 & Common Objects & Total \\
\hline
$<$4 & 0 & 3 & 0 & 3 \\
4-5 & 14 & 28 & 13 & 29 \\
5-6 & 62 & 140 & 57 & 145 \\
6-7 & 195 & 490 & 169 & 516 \\
7-8 & 675 & 1626 & 523 & 1778 \\
8-9 & 3108 & 4999 & 2099 & 6008 \\
\hline
\end{tabular}
\end{table}

\section{Low Resolution Spectroscopy}

\subsection{Observations}

Spectra were taken for a group of objects to give an indication of how effective the selection process weeded out giant contamination and to gain statistics on spectral type, temperature and metalicity of the M dwarf candidates.  Candidates were chosen solely based on whether they had observationally favourable locations on the sky at the time and place of the observations.  

One set of low resolution spectra were obtained using the Telescopio Nazionale Galileo (TNG) telescope on La Palma, Spain in May 2011.  These spectra were taken using the Device Optimized for the LOw RESolution (DOLORES) instrument using the LR-R grism.  The resolution of these spectra was approximately R$\sim$700.  Flux and wavelength calibration was performed using spectrophotometric standard stars and the Ne+Hg standard lamps.  The reduction process was carried out using IRAF routines.  Flats were median combined, bias subtracted, trimmed to remove any edge effects, and applied to the object images.  The spectra were extracted using IRAF.APALL which identified the apertures as well as calculated and subtracted the background.  The spectra of the spectrophotometric standard stars were used with the package IRAF.TELLURIC to determine the wavelength shift and scaling needed to account for the telluric lines in the atmosphere.  

Spectra were also collected in September 2011 from Ritchey-Chretien Focus Spectro-graph at the 4-m telescope in Kitt Peak, Arizona (KPNO) using the BL-181 grism.  The standard IRAF routines were also used for sky subtraction, wavelength and flux calibration employing the use of ThAr lamps and spectrophotometric standards.  After calibration and trimming, both sets of spectra cover a wavelength range of 6000-8000 \AA.  The KPNO spectra had a resolution of R$\sim$1000.	    
	
\subsection{Analysis}
	An initial spectral type was determined by comparing each of the observed spectra to that of known M dwarf standards from \cite{kirkpatrick_1} as well as M giant spectra from \cite{garcia}.  The standard spectra included K7-M9 dwarfs and K4-M4.5 giants.  The observed spectra were normalized using the mean value between 7450 and 7550 $\AA$.  The comparison standard stars were linearly interpolated so their resolution matched that of the observed spectra and normalized using the same region.   A least squares minimization was then performed to find the standard spectra that was the best fit to the observed spectra.  As well as providing a preliminary spectral type, this also provided an initial temperature estimate for later model fitting.  
	
Further refinement of spectral type and determination of whether the objects were giants was made by using the calcium hydride (CaH3) and titanium oxide (TiO) molecular bands.  The CaH3 (6960-6990\AA) region has been shown to display weaker absorption in giants than in dwarfs \citep{allen} and the full depth of the TiO5 (7126-7135\AA) feature has been shown to also be a good indication of spectral types for early M dwarfs \citep{kirkpatrick1993,PMSU}. 

\begin{figure}
\includegraphics[scale=.3]{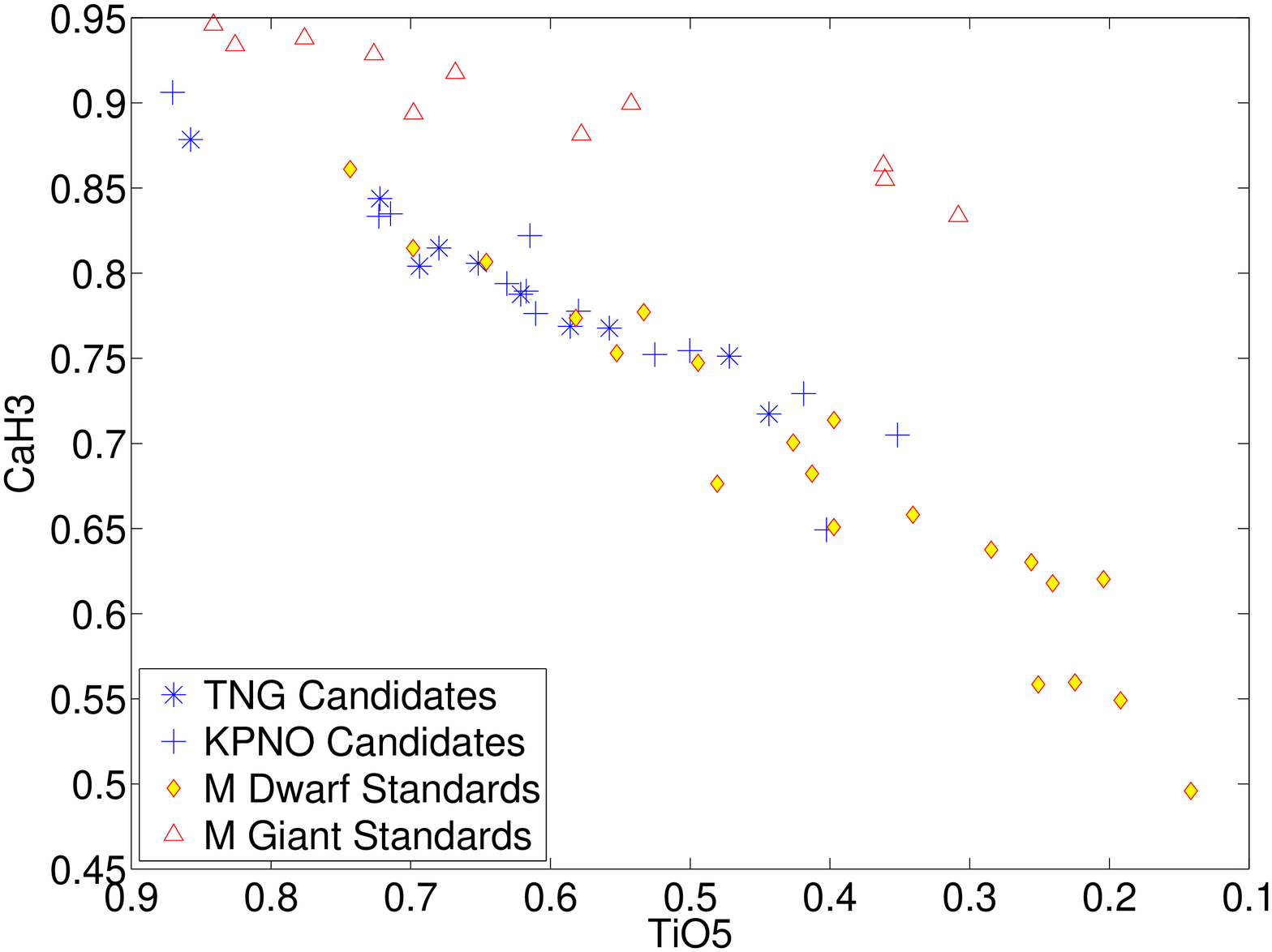}
	\centering
	\caption{CaH3 and TiO5 values for the observed targets.  M dwarf and M giant standards are plotted here as yellow diamonds and red triangles, respectfully, and show a vertical separation.  None of the candidate spectra fall within the region of the M giant standards indicating they are all likely to be M dwarfs.}
	\label{fig:cah3tio5}
\end{figure}

Since the CaH3 region is sensitive to gravity, when plotted together with the TiO5 bandstrenghth, any giants should stand out as clear outliers.  Figure \ref{fig:cah3tio5} shows the observed spectra in such a plot along with CaH3/TiO5 measurements taken of the M standards interpolated to the same resolution. The M dwarf and M giant standards show a vertical separation as expected with our observed spectra falling in the region consistent with M and K dwarfs.  Though this analysis only covers a small selection of objects within our sample, it is still encouraging that no giants were detected.

\begin{figure*}
\centering
\includegraphics[scale=.4]{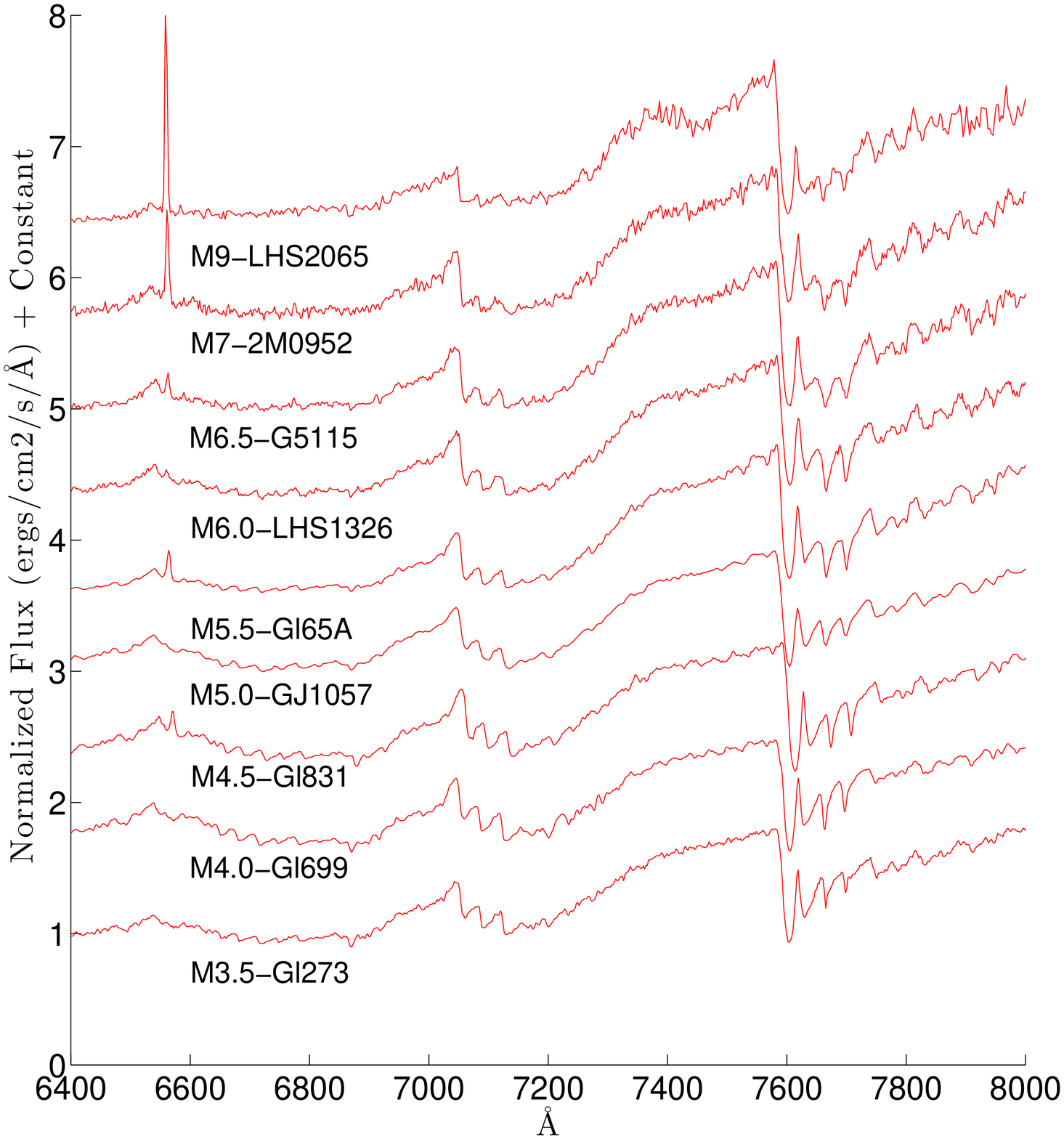}
\includegraphics[scale=.4]{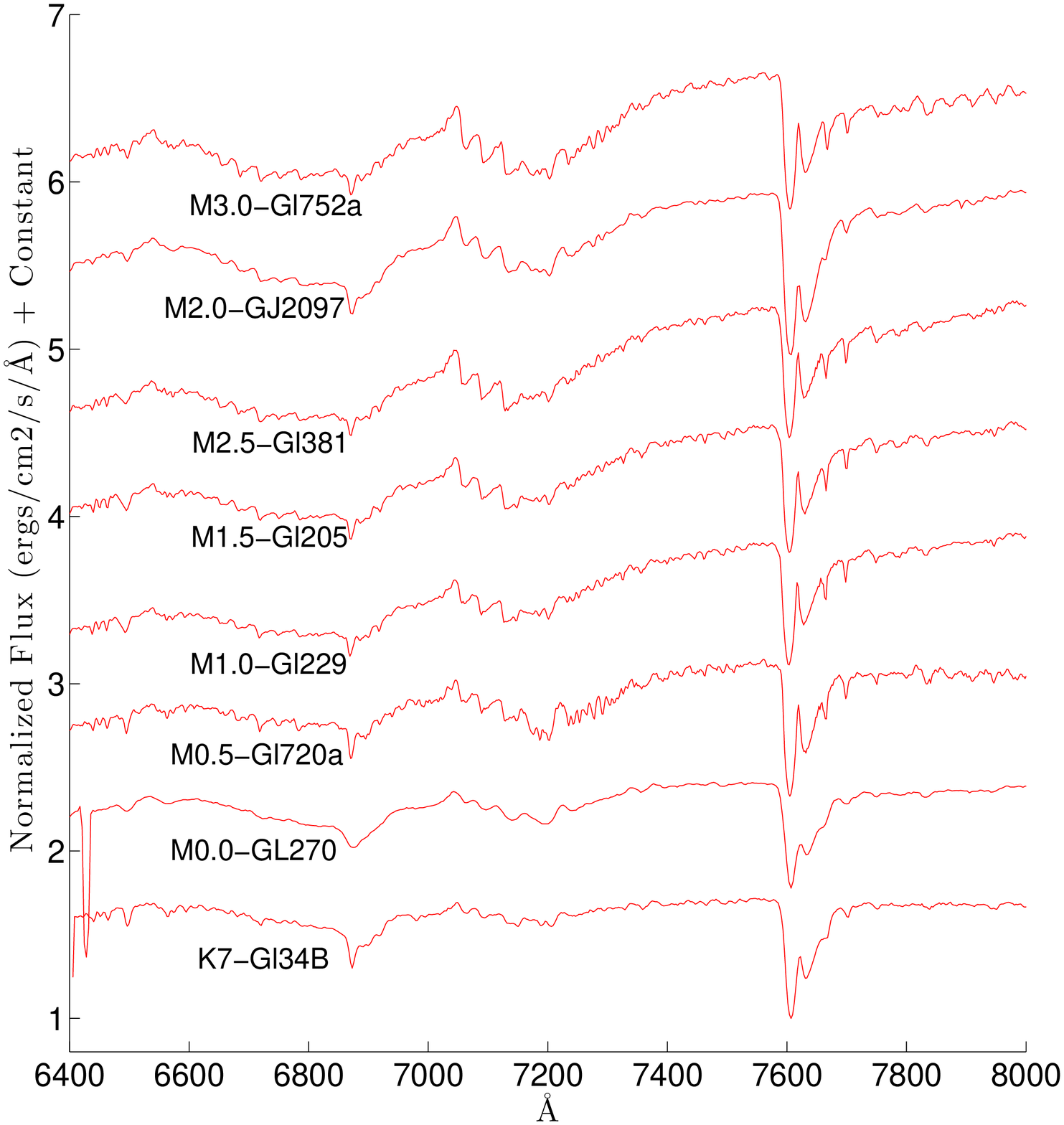}
\caption{M dwarfs used to determine spectral types of our targets.}
\label{fig:standards}
\end{figure*}

\begin{figure*}
\centering
\includegraphics[scale=.4]{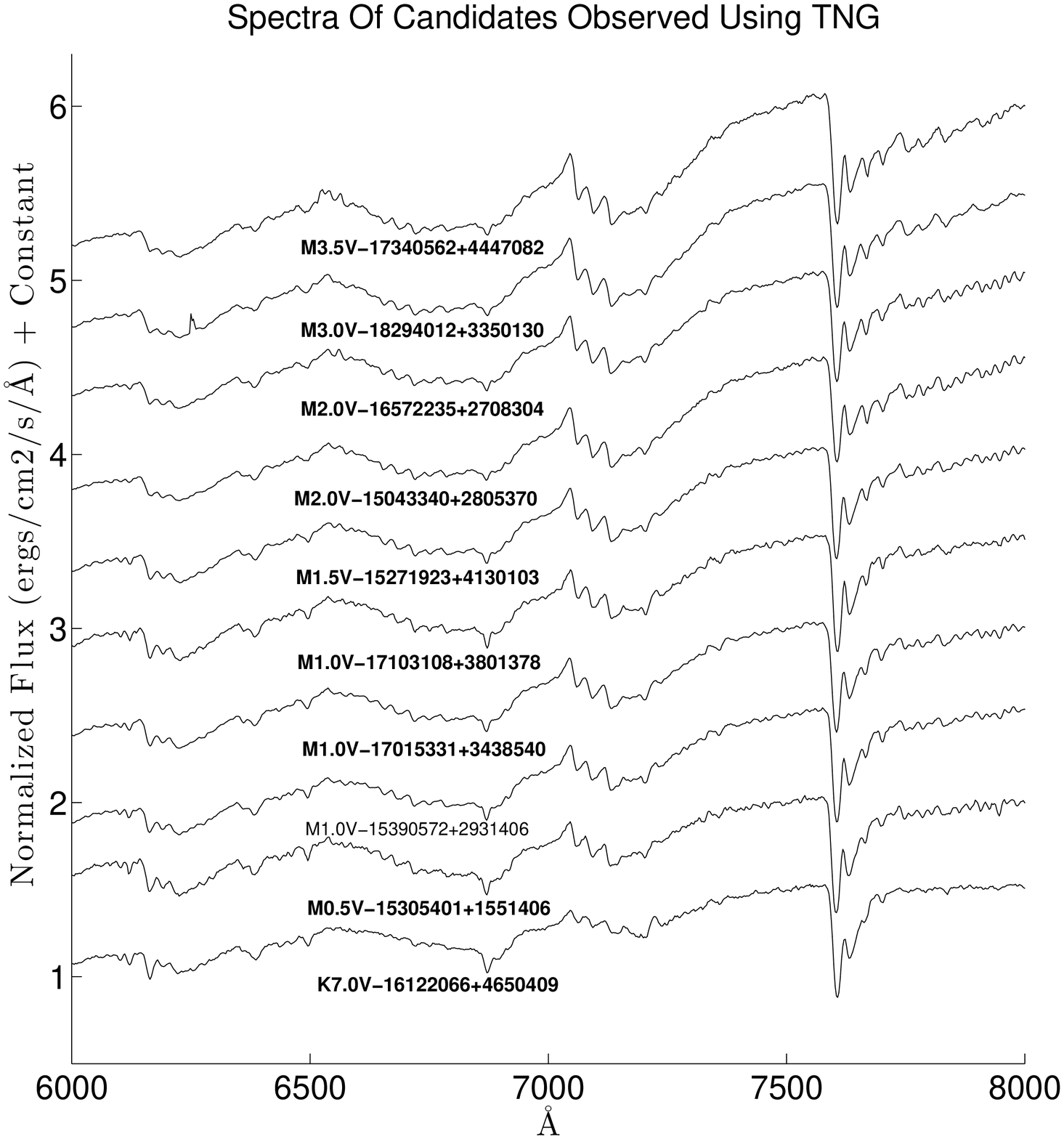}
\includegraphics[scale=.4]{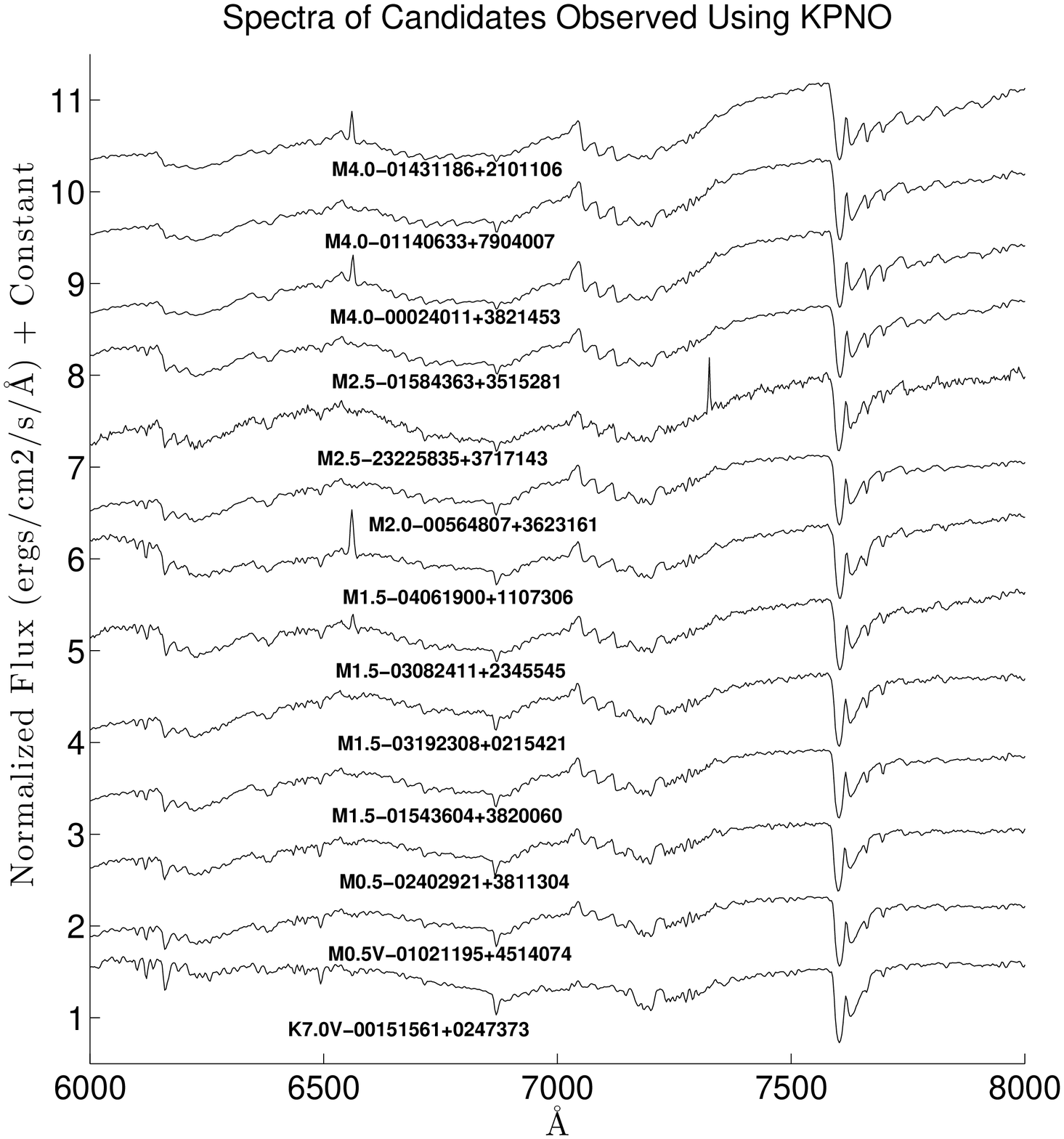}
\caption{All observed spectra of M dwarf candidates.  The spectra are labelled with our determined spectral type and 2MASS designation.}
\label{fig:obsspect}
\end{figure*}

We used the relationship between the TiO band strength and spectral type, derived by Ried et al, as an independent check of our own best fit spectral types.  Ried et al found this linear relationship to be
\begin{align*}
	S_p = -10.775\times TiO5+8.2
\end{align*}
Our spectral type fits agreed to within 0.5 spectral types of our least squares fit which is the stated uncertainty found by Ried et al with the above relationship.  The resulting spectral types can be found in Table \ref{tab:spectrainfo} and all of the observed spectra along with the M dwarf standard spectra used can be seen in Figures \ref{fig:standards} and \ref{fig:obsspect}.

Using the initial temperature estimations found in the spectral template least square fits above, a grid of synthetic spectra were generated using local thermal equilibrium  models from the NextGen code of \cite{hauschildt} by the program WITA6 \citep{wita}.  Opacity line lists for TiO were taken from \cite{plez} and other opacity sources are cited in \cite{yp_line_list_opacities}. A range of effective temperatures, metallicities, and gravities were generated at increments of 100 K, 0.5 dex, and 0.5 log $g$.  The synthetic spectra were created for 2600 K$<$ T $<$ 4600 K, 4.0 $<$ log $g$ $<$ 5.5, $-0.5$ $<$ [Fe/H] $<$ 0.5. We used the fitting procedure described in \cite{pavfit} and \cite{pavfit1} across 6000 - 8000\AA.  Only effective temperature's were iterated for the initial fit.  Metallicities and gravities were fixed such that [Fe/H] = 0.0 and a log $g$ = 5.0.  After the best fit for temperature was determined, [Fe/H] and log $g$ were iterated until the fitting parameter was minimized.  The results from the fits can be seen in Table \ref{tab:spectrainfo}.

\begin{table*}
\begin{minipage}[tbh]{1\linewidth}
\centering
\caption{Results of physical parameter fits for a selection of M dwarfs from this work.}
\label{tab:spectrainfo}
\centering
\begin{tabular}{ccccccccc}
\hline
 $2MASS Designation$ & $\alpha$ & $\delta$ & $K$ & $J-K$ & $SpTyp$ & $Teff$\footnote{Uncertainties for fitted stellar parameters: Teff $\pm$100k, log g $\pm$0.5,$[Fe/H]\pm$0.5} & log g & $[Fe/H]$\\
 & ($^{\circ}$) & ($^{\circ}$) & (mag) & (mag) & & (k) &(cgs)&(dex) \\
\hline
15305401+1551406  & 232.725081 & 15.861243 & 8.97 & 0.843 &M 0.5 V & 3800 & 5.0 & 0.0 \\
15043340+2805370  & 226.139194 & 28.093640 & 8.98 & 0.836 &M 2.0 V & 3300 & 4.5 & +0.5 \\
15390572+2931406  & 234.773838 & 29.527944 & 8.94 & 0.817 &M 1.0 V & 3600 & 5.0 & 0.0 \\
15271923+4130103  & 231.830081 & 41.502949 & 7.52 & 0.843 & M 1.5 V & 3600 & 5.0 & 0.0 \\
16572235+2708304  & 254.343100 & 27.141781 & 8.51 & 0.845 &M 2.0 V & 3500 & 5.0 & 0.0 \\
17015331+3438540  & 255.472140 & 34.64834 5 & 8.80 & 0.966 &M 1.0 V & 3500 & 4.5 & +0.5 \\
16122066+4650409  & 243.086139 & 46.844694 & 8.20 & 0.761 & K 7.0 V & 3900 & 5.0 & 0.0 \\
17103108+3801378  & 257.629549 & 38.027209 & 8.62 & 0.864 &M 1.0 V & 3700 & 5.0 & 0.0 \\
17340562+4447082  & 263.523338 & 44.785632 & 7.91 & 0.837 &M 3.5 V & 3300 & 5.0 & 0.0 \\
18294012+3350130  & 277.417172 & 33.836978 & 8.95 & 0.892 &M 3.0 V & 3400 & 5.0 & 0.0 \\
23225835+3717143  & 350.743198 & 37.287263 & 7.96 & 0.839 &M 2.5 V & 3600 & 5.0 & 0.0 \\
00024011+3821453  & 0.667193 & 38.362543 & 8.91 & 0.796 & M 4.0 V & 3400 & 5.0 & 0.0 \\
00564807+3623161  & 14.200329 & 36.387792 & 8.92 & 0.837 & M 2.0 V & 3700 & 5.0 & 0.0 \\
01021195+4514074  & 15.549841 & 45.235351 & 8.85 & 0.892 & M 0.5 V & 4000 & 5.0 & 0.0 \\
01140633+7904007  & 18.526638 & 79.066892 & 8.81 & 0.856 & M 3.0 V & 3500 & 5.0 & 0.0 \\
01543604+3820060  & 28.650189 & 38.334976 & 8.97 & 0.784 & M 1.5 V &3800 & 5.0 & 0.0 \\
02402921+3811304  & 40.121777 & 38.191774 & 8.78 & 0.815 & M 0.5 V & 3900 & 5.0 & 0.0 \\
03192308+0215421  & 49.846142 & 2.261704 & 8.99 & 0.839 & M 1.5 V & 3800 & 5.0 & 0.0 \\
00151561+0247373  & 3.815086 & 2.793711 & 8.66 & 0.867 & K 7.0 V & 4200 & 5.0 & 0.0 \\
01431186+2101106  & 25.799383 & 21.019610 & 8.36 & 0.890 & M 4.0 V & 3200 & 5.0 & 0.0 \\
01584363+3515281  & 29.681783 & 35.257797 & 8.90 & 0.753 & M 2.5 V & 3500 & 5.0 & 0.0 \\
03082411+2345545  & 47.100472 & 23.765137 & 8.85 & 0.860 & M 1.5 V & 3300 & 5.0 & 0.0 \\
04061900+1107306  & 61.579243 & 11.125108 & 8.91 & 0.836 & M 1.5 V & 3800 & 5.0 & 0.0 \\
\hline
\end{tabular}
\end{minipage}
\end{table*}

\section{Summary and Future Work}

This work has classified 4054 M dwarfs with magnitudes of K$<9$ from the PPMXL catalogue. By probing down to lower proper motions, this work has produced 1193 new bright M dwarf candidates that were not included in previous catalogues.  By combining these objects with M dwarfs from LG11, we present a catalogue with 8479 K$<$9 late K and M dwarfs for future transit studies.  

Good progress in transit discovery can be made by targeting this expansive M dwarf catalogue for low and high resolution spectroscopy.  Low resolution spectra will confirm the cool dwarf nature and high resolutions spectra will provide constraints on inclination angles of potential transiting planetary systems using the method outlined in \cite{Inclination_Herrero}.  Such observations would allow prioritization of bright M dwarfs for light curve follow-up and transit searches.

Target lists can then be expanded or refined for facilities such as Mearth \citep{Mearth_org}, Apache \citep{Apache}, and LCOGTN \citep{lcogt} as the search for transiting Super Earth's moves forward.

\section{Acknowledgements}

DJP, HRAJ, YP, JRB, and ELM have received support from and JF, MKK, MC, and RT were supported by RoPACS,  a Marie Curie Initial Training Network funded by the European Commission’s Seventh Framework Programme.  This research has benefitted from the M, L and T dwarf compendium housed at DwarfArchives.org and maintained by Chris Gelino, Davy Kirkpatrick and Adam Burgasser.  This research is based on observations made with the Italian Telescopio Nazionale Galileo (TNG) operated on the island of La Palma by the Fundación Galileo Galilei of the INAF (Istituto Nazionale di Astrofisica) at the Spanish Observatorio del Roque de los Muchachos of the Instituto de Astrofisica de Canarias and the Kitt Peak National Observatory which is operated by the Association of Universities for Research in Astronomy (AURA) under cooperative agreement with the National Science Foundation.

\label{lastpage}

\bibliographystyle{mn2e} 
\bibliography{mdwarf}

\end{document}